\documentclass[letterpaper,onecolumn,twoside,12pt]{IEEEtran}
\usepackage{amsmath, amssymb, cite, epsfig, psfrag}

\def\beq{\begin{equation}}
\def\eeq{\end{equation}}
\def\beqa{\begin{eqnarray}}
\def\eeqa{\end{eqnarray}}
\def\beqan{\begin{eqnarray*}}
\def\eeqan{\end{eqnarray*}}

\def\R{{\mathbb{R}}}

\def\argmax{\mathop{\mathrm{arg\,max}}}
\def\span{\mathop{\mathrm{span}}}
\def\x{\times}

\newtheorem{theorem}{Theorem}
\newtheorem{lemma}{Lemma}

\def\xhat{\hat{x}}
\def\Ihat{\ensuremath{\hat{I}}}
\def\IhatML{\ensuremath{\hat{I}_{\rm ML}}}
\def\IhatCorr{\ensuremath{\hat{I}_{\rm MC}}}
\def\DeltaAdd{\ensuremath{\Delta^+}}
\def\DeltaSub{\ensuremath{\Delta^-}}
\def\Itrue{\ensuremath{I_{\rm true}}}
\def\PMD{\ensuremath{p_{\rm MD}}}
\def\PFA{\ensuremath{p_{\rm FA}}}
\def\SNR{\mbox{\small \sffamily SNR}}
\def\MAR{\mbox{\small \sffamily MAR}}
\def\tableSNR{\mbox{\tiny \sffamily SNR}}
\def\tableMAR{\mbox{\tiny \sffamily MAR}}
\def\captionSNR{\mbox{\scriptsize \sffamily SNR}}
\def\captionMAR{\mbox{\scriptsize \sffamily MAR}}
\def\arr{\rightarrow}
\def\Exp{\mathbf{E}}
\newcommand{\Mover}{\ensuremath{\overline{M}}}
\newcommand{\Munder}{\ensuremath{\underline{M}}}
\newcommand{\BetaDist}{\mbox{Beta}}

\title{Necessary and Sufficient Conditions on \\ Sparsity Pattern Recovery}
\author{Alyson K. Fletcher,~\IEEEmembership{Member,~IEEE,}
        Sundeep Rangan,
        and~Vivek~K~Goyal,~\IEEEmembership{Senior~Member,~IEEE}%
\thanks{This work was supported in part by
	a University of California President's Postdoctoral Fellowship,
        NSF CAREER Grant CCF-643836, and
        the Centre Bernoulli at 
        \'{E}cole Polytechnique F\'{e}d\'{e}rale de Lausanne.}%
\thanks{A. K. Fletcher (email: alyson@eecs.berkeley.edu) is with
        the Department of Electrical Engineering and Computer Sciences,
        University of California, Berkeley.}
\thanks{S. Rangan (email: srangan@qualcomm.com) is with Qualcomm Technologies,
        Bedminster, NJ.}%
\thanks{V.~K. Goyal (email: vgoyal@mit.edu) is with
        the Department of Electrical Engineering and Computer Science and
        the Research Laboratory of Electronics,
        Massachusetts Institute of Technology.}}

\begin{document}

\maketitle

\begin{abstract}
The problem of detecting the sparsity pattern of a $k$-sparse vector
in $\R^n$ from $m$ random noisy measurements is of interest in
many areas such as system identification, denoising, pattern recognition,
and compressed sensing.
This paper addresses the scaling of the number of measurements $m$,
with signal dimension $n$ and sparsity-level nonzeros $k$,
for asymptotically-reliable detection. 
We show a necessary condition for perfect recovery
at any given SNR for all algorithms, regardless of complexity, 
is $m = \Omega(k\log(n-k))$ measurements.
Conversely, it is shown that this scaling of $\Omega(k\log(n-k))$ 
measurements is sufficient
for a remarkably simple ``maximum correlation'' estimator.
Hence this scaling is optimal and does not require more sophisticated
techniques such as lasso or matching pursuit.
The constants for both the necessary and sufficient conditions are
precisely defined in terms of the minimum-to-average ratio of the
nonzero components and the SNR\@.
The necessary condition improves upon 
previous results for maximum likelihood estimation.  
For lasso, it also provides a necessary condition 
at any SNR and for low SNR improves upon previous work.
The sufficient condition provides the first 
asymptotically-reliable detection guarantee
at finite SNR\@.
\end{abstract}


\section{Introduction}
Suppose one is given an observation $y \in \R^m$ that was generated
through $y = A x + d$, where $A \in \R^{m \x n}$ is known and
$d \in \R^m$ is an additive noise vector with a known distribution.
It may be desirable for an estimate of $x$ to have a
small number of nonzero components.
An intuitive example is when one wants to choose a small subset from
a large number of possibly-related factors that linearly influence
a vector of observed data.
Each factor corresponds to a column of $A$, 
and one wishes to find a small subset of columns with which to
form a linear combination that closely matches the observed data $y$.
This is the subset selection problem in (linear) regression~\cite{Miller:02},
and it gives no reason to penalize large values for the nonzero components.

In this paper, we assume that the true signal $x$ has $k$ nonzero entries
and that $k$ is known when estimating $x$ from $y$.
We are concerned with establishing necessary and sufficient conditions
for the recovery of the \emph{positions} of the nonzero entries of $x$,
which we call the \emph{sparsity pattern}.
Once the sparsity pattern is correct, $n-k$ columns of $A$ can be
ignored and the stability of the solution is well understood;
however, we do not study any other performance criterion.

\subsection{Previous Work}
Sparsity pattern recovery (or more simply, sparsity recovery)
has received considerable attention in a
variety guises.  Most transparent from our formulation is the connection
to sparse approximation.  In a typical sparse approximation problem,
one is given data $y \in \R^m$, dictionary\footnote{The term seems
to have originated in~\cite{MallatZ:93} and may apply to $A$ or the
columns of $A$ as a set.} $A \in \R^{m \x n}$, and tolerance $\epsilon > 0$.
The aim is to find $\xhat$ with the fewest number of nonzero entries
among those satisfying $\| A \xhat - y \| \leq \epsilon$.
This problem is NP-hard~\cite{Natarajan:95} but greedy heuristics
(matching pursuit~\cite{MallatZ:93} and its variants) and
convex relaxations (basis pursuit~\cite{ChenDS:99}, lasso~\cite{Tibshirani:96}
and others) can be effective under certain conditions on
$A$ and $y$~\cite{DonohoET:06,Tropp:04,Tropp:06}.
Scaling laws for sparsity recovery with any $A$ were first given
in~\cite{FletcherRGR:06}.

More recently, the concept of ``sensing'' sparse $x$ through multiplication
by a suitable random matrix $A$, with measurement error $d$, has been
termed \emph{compressed sensing}~\cite{CandesRT:06-IT,Donoho:06,CandesT:06}.
This has popularized the study of sparse approximation with respect
to random dictionaries, which was considered also in~\cite{FletcherRG:05}.
Results are generally phrased as the asymptotic scaling of the
number of measurements $m$ (the length of $y$)
needed for sparsity recovery to succeed with high probability,
as a function of the other problem parameters.
More specifically, most results are sufficient conditions for
specific tractable recovery algorithms to succeed.
For example, if $A$ has i.i.d.\ Gaussian entries and $d = 0$,
then $m \asymp 2 k \log(n/k)$ dictates the minimum scaling at which
basis pursuit succeeds with high probability~\cite{DonohoT:0x}.
With nonzero noise variance,
necessary and sufficient conditions for the success of lasso in this
setting have the asymptotic scaling~\cite{Wainwright:06}
\beq
 \label{eq:WainwrightLasso}
  m \asymp 2 k \log(n-k) + k + 1.
\eeq

To understand the ultimate limits of sparsity recovery, while also
casting light on the efficacy of lasso or orthogonal matching pursuit (OMP),
it is of interest to determine necessary and sufficient conditions for
an optimal recovery algorithm to succeed.
Of course, since it is sufficient for lasso,
the condition (\ref{eq:WainwrightLasso}) is sufficient
for an optimal algorithm.  Is it close to a necessary condition?
We address precisely this question by proving a necessary condition
that differs from (\ref{eq:WainwrightLasso})
by a factor that is \emph{constant with respect to $n$ and $k$}
while depending on the signal-to-noise ratio (SNR) and
mean-to-average ratio (MAR),
which will be defined precisely in Section~\ref{sec:Statement}.
Furthermore, we present an extremely simple algorithm for which
a sufficient condition for sparsity recovery is similarly within a constant
factor of (\ref{eq:WainwrightLasso}).

Previous necessary conditions had been
based on information-theoretic analyses such as the
capacity arguments in~\cite{SarvothamBB:06-Allerton,FletcherRG:07b}
and a use of Fano's inequality in~\cite{Wainwright:07-725}.
More recent publications with necessary conditions
include~\cite{GoyalFR:08,Reeves:08,AkcakayaT:07arXiv,AkcakayaT:CISS08}.
As described in Section~\ref{sec:Necessary}, our new necessary conditions
are stronger than the previous results.

Table~\ref{table:summary} previews our main results and places
(\ref{eq:WainwrightLasso}) in context.
The measurement model and parameters $\MAR$ and $\SNR$
are defined in the following section.
Arbitrarily small constants have been omitted, and
the last column---labeled simply $\SNR \rightarrow \infty$---is
more specifically for $\MAR > \epsilon > 0$ for some fixed $\epsilon$
and $\SNR = \Omega(k)$.

\begin{table*}
 \begin{center}
  \begin{tabular}{|c||c|c|}
    \hline
           & finite $\captionSNR$ & $\captionSNR \rightarrow \infty$ \\
\hline \hline
 & & \\[-2mm]
 Any algorithm must fail & $m < \frac{2}
                          {\tableMAR \cdot \tableSNR}k \log(n-k) + k - 1$
           & $m \leq k$ \\
           & Theorem~\ref{thm:minMeasML}
           & (elementary) 
 \\[-2mm]
 & & \\ \hline
 & & \\[-2mm]
 Necessary and       
           & unknown (expressions above    & $m \asymp 2 k \log(n-k) + k + 1$ \\
 sufficient for lasso
           & and right are necessary)      & Wainwright~\cite{Wainwright:06}
 \\[-2mm]
 & & \\ \hline
 & & \\[-2mm]
 Sufficient for maximum 
           & $m > \frac{8(1+\tableSNR)}{\tableMAR \cdot \tableSNR}k \log(n-k)$
           & $m > \frac{8}{\tableMAR}k \log(n-k)$ \\
 correlation estimator (\ref{eq:corrEst})
           & Theorem~\ref{thm:corrBnd}
           & from Theorem~\ref{thm:corrBnd}
 \\[-2mm]
 & & \\ \hline
  \end{tabular}
 \end{center}
 \caption{Summary of Results on Measurement Scaling for
   Reliable Sparsity Recovery \newline
   (see body for definitions and technical limitations)}
 \label{table:summary}
\end{table*}

\subsection{Paper Organization}
The setting
is formalized in Section~\ref{sec:Statement}.
In particular, we define our concepts of signal-to-noise ratio
and mean-to-average ratio; our results clarify the roles of these
quantities in the sparsity recovery problem.
Necessary conditions for success of any algorithm are considered
in Section~\ref{sec:Necessary}.  There we present a new necessary
condition and compare it to previous results and numerical experiments.
Section~\ref{sec:Sufficient} introduces a very simple recovery
algorithm for the purpose of showing that a sufficient condition
for its success is rather weak---it has the same dependence on
$n$ and $k$ as (\ref{eq:WainwrightLasso}).
Conclusions are given in Section~\ref{sec:Conclusions},
and proofs appear in the Appendix.

\section{Problem Statement}
\label{sec:Statement}
Consider estimating a $k$-sparse vector
$x \in \R^n$ through a vector of observations,
\beq
    y = A x + d,
\eeq
where $A \in \R^{m \x n}$ is a random matrix with i.i.d.\
${\mathcal{N}}(0,1/m)$ entries
and $d \in \R^m$ is i.i.d.\ unit-variance Gaussian noise.
Denote the sparsity pattern of $x$ (positions of nonzero entries)
by the set $\Itrue$,
which is a $k$-element subset of the set of indices $\{1,\,2,\,\ldots,\,n\}$.
Estimates of the sparsity pattern will be denoted by $\Ihat$
with subscripts indicating the type of estimator.
We seek conditions under which there exists an estimator
such that $\Ihat = \Itrue$ with high probability.

In addition to the signal dimensions, $m$, $n$ and $k$, we will show
that there are two variables that dictate the ability to
detect the sparsity pattern reliably:  the SNR, and what we will call the
\emph{minimum-to-average ratio}.

The SNR is defined by
\beq \label{eq:snrDef}
    \SNR = \frac{\Exp[\|Ax\|^2]}{\Exp[\|d\|^2]}
         = \frac{\Exp[\|Ax\|^2]}{m}.
\eeq
Since we are considering $x$ as an unknown deterministic vector,
the SNR can be further simplified as follows:
The entries of $A$ are i.i.d.\ ${\mathcal{N}}(0,1/m)$,
so columns $a_i \in \R^m$ and $a_j \in \R^m$ of $A$ satisfy
$\Exp[a_i' a_j] = \delta_{ij}$.
Therefore, the signal energy is given by
\beqan
  \Exp\left[\|Ax\|^2\right]
  & = & \Exp\left[ \| {\sum_{j \in \Itrue} a_jx_j} \|^2 \right]\\
  & = & \mathop{\sum\sum}_{i,j \in \Itrue} \Exp\left[ a_i' a_j x_i x_j \right] \\
  & = & \mathop{\sum\sum}_{i,j \in \Itrue} x_i x_j \delta_{ij}
  %  =   \Exp\left[ \sum_{j \in \Itrue}\|a_j\|^2|x_j|^2 \right]
  \ = \ \|x\|^2.
\eeqan
Substituting into the definition (\ref{eq:snrDef}), the SNR is given by
\beq \label{eq:snrVal}
    \SNR = \frac{1}{m}\|x\|^2.
\eeq

The minimum-to-average ratio of $x$ is defined as
\beq \label{eq:MAR-def}
    \MAR = \frac{\min_{j \in \Itrue} |x_j|^2}
                {\|x\|^2/k} .
\eeq
Since $\|x\|^2 / k$ is the average of $\{ |x_j|^2 \mid j \in \Itrue\}$,
$\MAR \in (0,1]$ with the upper limit occurring when all the nonzero
entries of $x$ have the same magnitude.

\subsubsection*{Remarks}
Other works use a variety of normalizations, e.g.:
the entries of $A$ have variance $1/n$ in~\cite{CandesT:06,Reeves:08};
the entries of $A$ have unit variance and the variance of $d$
is a variable $\sigma^2$
in~\cite{Wainwright:06,Wainwright:07-725,AkcakayaT:07arXiv,AkcakayaT:CISS08};
and our scaling of $A$ and a noise variance of $\sigma^2$ are used
in~\cite{HauptN:06}.
This necessitates great care in comparing results.

Some results involve
$$
  \MAR \cdot \SNR = \frac{k}{m} \min_{j \in \Itrue} |x_j|^2.
$$
While a similar quantity affects a regularization weight sequence
in~\cite{Wainwright:06}, there it does not affect the number of
measurements required for the success of lasso.\footnote{The formulation
of~\cite{Wainwright:06} makes $\captionSNR = \Theta(n)$, which obscures the
effect of the noise level.  See also the second remark following
Theorem~\ref{thm:corrBnd}.}
The magnitude of the smallest nonzero entry of $x$ is also prominent
in the phrasing of results in~\cite{AkcakayaT:07arXiv,AkcakayaT:CISS08}.

\section{Necessary Condition for Sparsity Recovery}
\label{sec:Necessary}
We first consider sparsity recovery without being concerned with
computational complexity of the estimation algorithm.
Since the vector $x \in \R^n$ is $k$-sparse,
the vector $Ax$ belongs to one of $L = {n \choose k}$ subspaces spanned by
$k$ of the $n$ columns of $A$.
Estimation of the sparsity pattern is the selection of one of these subspaces,
and since the noise $d$ is Gaussian,
the probability of error is minimized by choosing the
subspace closest to the observed vector $y$.
This results in the maximum likelihood (ML) estimate.

Mathematically, the ML estimator can be described as follows.
Given a subset $J \subseteq \{1,\,2,\,\ldots,\,n\}$,
let $P_J y$ denote the orthogonal projection of the vector $y$
onto the subspace spanned by the vectors $\{a_j \mid j \in J\}$.
The ML estimate of the sparsity pattern is
\[
    \IhatML = \argmax_{J~:~|J|=k} \|P_Jy\|^2,
\]
where $|J|$ denotes the cardinality of $J$.
That is, the ML estimate is the set of $k$
indices such that the subspace spanned by the corresponding columns of $A$
contain the maximum signal energy of $y$.

Since the number of subspaces, $L$, grows exponentially in $n$ and $k$,
an exhaustive search is computationally infeasible.  However, the performance
of ML estimation provides a lower bound on the number of measurements
needed by any algorithm
that cannot exploit a priori information on $x$ other than
it being $k$-sparse.

\begin{theorem}
\label{thm:minMeasML}
Let $k = k(n)$ and $m = m(n)$ vary with $n$ such that
$\lim_{n\arr\infty} k(n) = \infty$
and
\beq \label{eq:minMeasML}
    m(n) < \frac{2 - \delta}{\MAR \cdot \SNR}k \log(n-k) + k - 1
\eeq
for some $\delta > 0$.
Then even the ML estimator asymptotically cannot detect the
sparsity pattern, i.e.,
\[
    \lim_{n \arr \infty} \Pr\left( \IhatML = \Itrue \right) = 0.
\]
\end{theorem}
\begin{proof}
See Appendix~\ref{sec:NecProof}.
\end{proof}

The theorem shows that for fixed {\SNR} and {\MAR},
the scaling $m = \Omega(k\log(n-k))$ is necessary for reliable sparsity
pattern recovery.
The next section will show that this scaling can be achieved with
an extremely simple method.

\subsubsection*{Remarks}
\begin{enumerate}
\item
The theorem applies for any $k(n)$ such that
$\lim_{n\rightarrow\infty} k(n) = \infty$,
including both cases with $k = o(n)$ and $k = \Theta(n)$.
In particular, under linear sparsity
($k = \alpha n$ for some constant $\alpha$),
the theorem shows that
$$
  m \asymp \frac{2\alpha}{\MAR \cdot \SNR} n \log n
$$
measurements are necessary for sparsity recovery.
Similarly, if $m/n$ is bounded above by a constant, then
sparsity recovery will certainly fail unless $k = O(n/\log n)$.
\item
In the case of $\MAR \cdot \SNR < 1$, the bound (\ref{eq:minMeasML})
improves upon the necessary condition of~\cite{Wainwright:06}
for the asymptotic success of lasso
by the factor $(\MAR \cdot \SNR)^{-1}$.
\item
The bound (\ref{eq:minMeasML}) can be compared against
the information-theoretic bounds mentioned earlier.
The tightest of these bounds is in \cite{FletcherRG:07b} and shows
that the problem dimensions must satisfy
\beq \label{eq:capBnd}
    \frac{2}{m}\log_2{n \choose k} \leq \log_2(1 + \SNR) - \alpha
    \log_2 (1 + \frac{\SNR}{\alpha}),
\eeq
where $\alpha = k/n$ is the \emph{sparsity ratio}.
For large $n$ and $k$, the bound can be rearranged as
\[
    m \geq
    \frac{2h(\alpha)}{\alpha}
    \left[ \log_2(1+\SNR) - \alpha \log_2(1+\frac{\SNR}{\alpha})\right]^{-1} k,
\]
where $h(\cdot)$ is the binary entropy function.
In particular, when the sparsity ratio $\alpha$ is fixed,
the bound shows only that $m$ needs to grow at least linearly with $k$.
In contrast, Theorem~\ref{thm:minMeasML} shows that with fixed sparsity ratio
$m = \Omega(k\log(n-k))$ is necessary for reliable sparsity recovery.
Thus, the bound in Theorem~\ref{thm:minMeasML} is significantly tighter
and reveals that the previous information-theoretic necessary conditions
from~\cite{SarvothamBB:06-Allerton,FletcherRG:07b,Wainwright:07-725,AkcakayaT:07arXiv,AkcakayaT:CISS08}
are overly optimistic.

\item
Results more similar to Theorem~\ref{thm:minMeasML}---based on direct analyses
of error events rather than information-theoretic arguments---appeared
in~\cite{GoyalFR:08,Reeves:08}.
The previous results showed that with fixed SNR as defined here,
sparsity recovery with $m = \Theta(k)$ must fail.
The more refined analysis in this paper gives the additional $\log(n-k)$
factor and the precise dependence on $\MAR \cdot \SNR$.

\item
Theorem~\ref{thm:minMeasML} is not contradicted by the relevant sufficient
condition of~\cite{AkcakayaT:CISS08,AkcakayaT:07arXiv}.
That sufficient condition holds for scaling that gives linear sparsity and 
$\MAR \cdot \SNR = \Omega(\sqrt{n \log n})$.
For $\MAR \cdot \SNR = \sqrt{n \log n}$, Theorem~\ref{thm:minMeasML} shows
that fewer than $m \asymp 2\sqrt{k \log k}$ measurements will cause
ML decoding to fail, while~\cite[Thm.~3.1]{AkcakayaT:CISS08} shows
that a typicality-based decoder will succeed with $m = \Theta(k)$ measurements.

\item
Note that the necessary condition of~\cite{Wainwright:07-725}
is proven for $\MAR = 1$.
Theorem~\ref{thm:minMeasML} gives a bound that increases for smaller $\MAR$;
this suggests (though does not prove, since the condition is merely necessary)
that smaller $\MAR$ makes the problem harder.
\end{enumerate}

\subsubsection*{Numerical validation}
Computational confirmation of Theorem~\ref{thm:minMeasML} is technically
impossible, and even qualitative support is hard to obtain because of
the high complexity of ML detection.
Nevertheless, we may obtain some evidence through Monte Carlo simulation.

\begin{figure*}
 \begin{center}
  \psfrag{k}[][]{$k$}
  \psfrag{m}[][]{$m$}
  \epsfig{figure=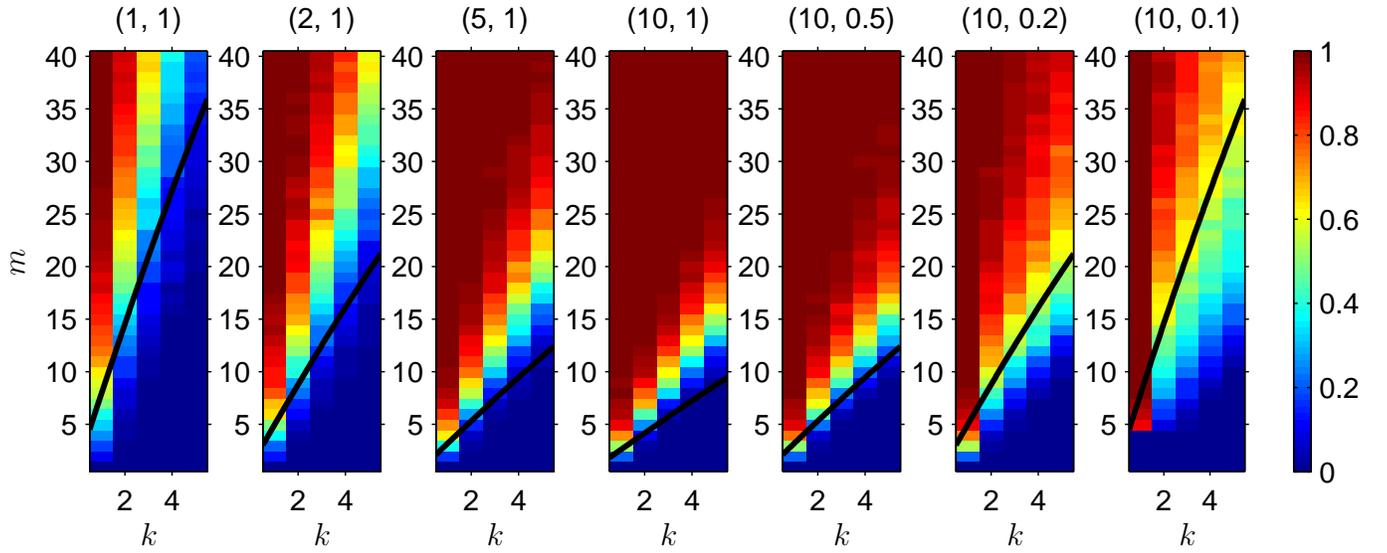,width=7.15in}
 \end{center}
 \caption{Simulated success probability of ML detection for $n=20$ and
   many values of $k$, $m$, $\captionSNR$, and $\captionMAR$.
   Each subfigure gives simulation results for $k \in \{1,2,\ldots,5\}$
   and $m \in \{1,2,\ldots,40\}$ for one $(\captionSNR,\captionMAR)$ pair.
   Each subfigure heading gives $(\captionSNR,\captionMAR)$.
   Each point represents at least 500 independent trials.
   Overlaid on the color-intensity plots
   is a black curve representing (\ref{eq:minMeasML}).}
 \label{fig:ML}
\end{figure*}

Fig.~\ref{fig:ML} shows the probability of success of ML detection for
$n = 20$ as $k$, $m$, $\SNR$, and $\MAR$ are varied, with each point
representing at least 500 independent trials.
Each subpanel gives simulation results for $k \in \{1,2,\ldots,5\}$
and $m \in \{1,2,\ldots,40\}$ for one $(\SNR,\MAR)$ pair.
Signals with $\MAR < 1$ are created by having one small nonzero component
and $k-1$ equal, larger nonzero components.
Overlaid on the color-intensity plots is a black curve representing
(\ref{eq:minMeasML}).

Taking any one column of one subpanel from bottom to top shows that
as $m$ is increased, there is a transition from ML failing to ML succeeding.
One can see that (\ref{eq:minMeasML}) follows the failure-success
transition qualitatively.
In particular, the empirical dependence on $\SNR$ and $\MAR$ approximately
follows (\ref{eq:minMeasML}).
Empirically, for the (small) value of $n = 20$,
it seems that with $\MAR \cdot \SNR$ held fixed,
sparsity recovery becomes easier as $\SNR$ increases
(and $\MAR$ decreases).

\begin{figure}
 \begin{center}
  \psfrag{k}[][]{$k$}
  \psfrag{m}[][]{$m$}
  \epsfig{figure=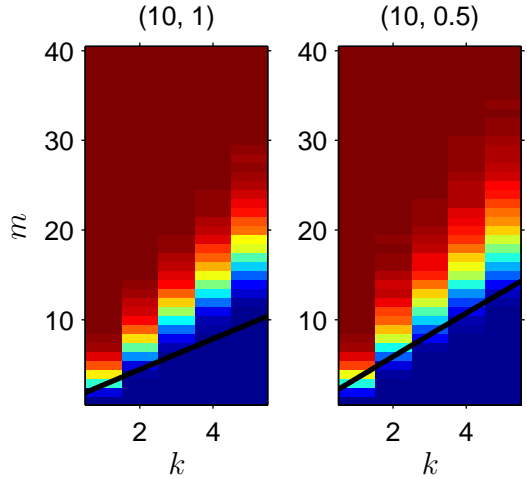,width=2.8in}
 \end{center}
 \caption{Simulated success probability of ML detection for $n=40$;
   $\captionSNR = 10$; $\captionMAR = 1$ (left) or $\captionMAR = 0.5$ (right);
   and many values of $k$ and $m$.
   Each subfigure gives simulation results for $k \in \{1,2,\ldots,5\}$
   and $m \in \{1,2,\ldots,40\}$, with
   each point representing at least 1000 independent trials.
   Overlaid on the color-intensity plots (with scale as in Fig.~\ref{fig:ML})
   is a black curve representing (\ref{eq:minMeasML}).}
 \label{fig:ML40}
\end{figure}

Less extensive Monte Carlo simulations for $n=40$ are reported
in Fig.~\ref{fig:ML40}.  The results are qualitatively similar.
As might be expected, the transition from low to high probability
of successful recovery as a function of $m$ appears more sharp
at $n=40$ than at $n=20$.

\section{Sufficient Condition with Maximum Correlation Detection}
\label{sec:Sufficient}
Consider the following simple estimator.
As before, let $a_j$ be the $j$th column of the random matrix $A$.
Define the \emph{maximum correlation (MC) estimate} as
\beq \label{eq:corrEst}
    \IhatCorr = \left\{ j ~:~ |a_j' y| \;
               \mbox{is one of the $k$ largest values of $|a_i' y|$} \right\}.
\eeq
This algorithm simply correlates the observed signal $y$
with all the frame vectors $a_j$ and selects the indices $j$
with the highest correlation.  It is significantly simpler than
both lasso and matching pursuit and is not meant to be proposed as
a competitive alternative.  Rather, the MC method is introduced and
analyzed to illustrate that a trivial method can obtain optimal
scaling with respect to $n$ and $k$.

\begin{theorem} \label{thm:corrBnd}
Let $k = k(n)$ and $m = m(n)$ vary with $n$ such that
$\lim_{n\arr\infty} k = \infty$,
$\limsup_{n\arr\infty} k/n \leq 1/2$,
and
\beq \label{eq:minMeasCorr}
    m > \frac{(8 + \delta)(1+\SNR)}{\MAR \cdot \SNR}k \log(n-k)
\eeq
for some $\delta > 0$.
Then the maximum correlation estimator
asymptotically detects the sparsity pattern,
i.e.,
\[
    \lim_{n \arr \infty} \Pr\left( \IhatCorr = \Itrue \right) = 1.
\]
\end{theorem}
\begin{proof}
See Appendix~\ref{sec:SuffProof}.
\end{proof}

\subsubsection*{Remarks}
\begin{enumerate}
\item
Comparing (\ref{eq:minMeasML}) and (\ref{eq:minMeasCorr}),
we see that for a fixed SNR and minimum-to-average ratio,
the simple MC estimator needs only a constant factor more measurements
than the optimal ML estimator.  In particular, the results show that
the scaling of the minimum number of measurements $m = \Theta(k\log(n-k))$
is both necessary and sufficient.  Moreover, the optimal
scaling factor not only does not require ML estimation,
it does not even require lasso or matching pursuit---it can be achieved
with a remarkably simply method such as maximum correlation.

\hspace*{1em}
There is, of course, a difference in the constant factors of the
expressions (\ref{eq:minMeasML}) and (\ref{eq:minMeasCorr}).
Specifically, the MC method requires a factor
    $4(1+\SNR)$
more measurements than ML detection.  In particular, for low SNRs
(i.e.\ $\SNR \ll 1$), the factor reduces to 4\@.

\item
For high SNRs, the gap between the MC estimator
and the ML estimator can be large.
In particular, the lower bound on the number of measurements
required by ML decreases to $k-1$ as $\SNR \rightarrow \infty$.\footnote{Of
course, at least $k+1$ measurements are necessary.}
In contrast, with the MC estimator
increasing the SNR has diminishing returns:
as $\SNR \arr \infty$, the bound on the
number of measurements in (\ref{eq:minMeasCorr}) approaches
\beq \label{eq:minMeasCorrLim}
    m > \frac{8}{\MAR}k\log(n-k).
\eeq
Thus, even with $\SNR \arr \infty$,
the minimum number of measurements is not improved from
$m = O(k\log(n-k))$.

\hspace*{1em}
This diminishing returns for improved SNR exhibited by the
MC method is also a problem for more sophisticated methods such as lasso.
For example, the analysis of~\cite{Wainwright:06}
shows that when the $\SNR = \Theta(n)$ (so $\SNR \arr \infty$) and
$\MAR$ is bounded strictly away from zero,
lasso requires
\beq \label{eq:minMeasLasso}
    m > 2k\log(n-k) + k + 1
\eeq
for reliable recovery.
Therefore, like the MC method,
lasso does not achieve a scaling better than $m = O(k\log(n-k))$,
even at infinite SNR\@.

\item
There is certainly a gap between MC and lasso.
Comparing (\ref{eq:minMeasCorrLim}) and (\ref{eq:minMeasLasso}),
we see that, at high SNR, the simple MC method requires
a factor of at most $4/\MAR$ more measurements than lasso.
This factor is largest when $\MAR$ is small,
which occurs when there are relatively small non-zero components.
Thus, in the high SNR regime,
the main benefit of lasso is not that it achieves an optimal scaling
with respect to $k$ and $n$
(which can be achieved with the simpler MC),
but rather that lasso is able to detect small coefficients,
even when they are much below the average power.
\end{enumerate}

\subsubsection*{Numerical validation}
\begin{figure*}
 \begin{center}
  \psfrag{K}[][]{$k$}
  \psfrag{M}[][]{$m$}
  \epsfig{figure=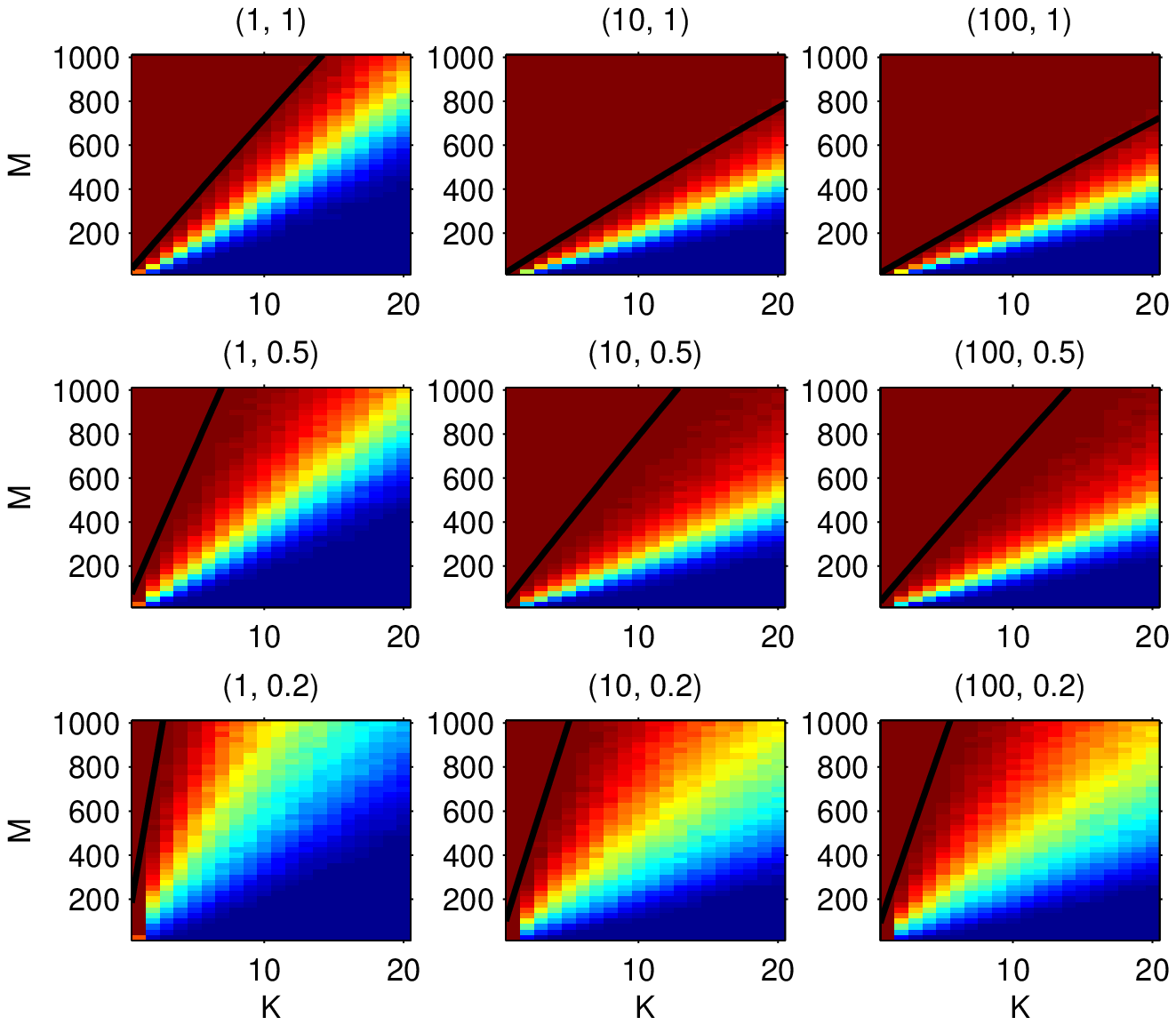,width=6in}
 \end{center}
 \caption{Simulated success probability of MC detection for $n=100$ and
   many values of $k$, $m$, $\captionSNR$, and $\captionMAR$.
   Each subfigure gives simulation results for $k \in \{1,2,\ldots,20\}$
   and $m \in \{25,50,\ldots,1000\}$ for one $(\captionSNR,\captionMAR)$ pair.
   Each subfigure heading gives $(\captionSNR,\captionMAR)$,
   so $\captionSNR = 1, 10, 100$ for the three columns and
   $\captionMAR = 1, 0.5, 0.2$ for the three rows.
   Each point represents 1000 independent trials.
   Overlaid on the color-intensity plots
   (with scale as in Fig.~\ref{fig:ML})
   is a black curve representing (\ref{eq:minMeasCorr}).}
 \label{fig:MC}
\end{figure*}

MC sparsity pattern detection is extremely simple and can thus be
simulated easily for large problem sizes.
Fig.~\ref{fig:MC} reports the results of a large number Monte Carlo simulations
of the MC method with $n=100$.
The threshold predicted by (\ref{eq:minMeasCorr})
matches well to the parameter combinations where
the probability of success drops below about 0.995\@.

\section{Conclusions}
\label{sec:Conclusions}
We have considered the problem of detecting the sparsity pattern
of a sparse vector from noisy random linear measurements.
Our main conclusions are:
\begin{itemize}
\item  \emph{Necessary and sufficient scaling with respect to $n$ and $k$.}
  For a given SNR and minimum-to-average ratio,
  the scaling of the number of measurements
  $$m = O(k\log(n-k))$$
  is both necessary and sufficient for asymptotically reliable sparsity
  pattern detection.
  This scaling is significantly worse than predicted by
  previous information-theoretic bounds.
\item  \emph{Scaling optimality of a trivial method.}
  The optimal scaling with respect to $k$ and $n$ can be achieved
  with a trivial maximum correlation (MC) method.
  In particular, both lasso and OMP, while likely to do better,
  are not necessary to achieve this scaling.
\item  \emph{Dependence on SNR.}
  While the threshold number of measurements for ML and MC
  sparsity recovery to be successful have the same dependence on
  $n$ and $k$, the dependence on SNR differs significantly.
  Specifically, the MC method requires a factor of up to $4(1+\SNR)$
  more measurements than ML\@.  Moreover, as $\SNR \arr \infty$,
  the number of measurements required by ML may be as low as $m = k+1$.
  In contrast, even letting $\SNR \arr \infty$,
  the maximum correlation method still requires a scaling $m = O(k\log(n-k))$.
\item  \emph{Lasso and dependence on MAR.}
  MC can also be compared to lasso, at least in the high SNR regime.
  There is a potential gap between MC and lasso,
  but the gap is smaller than the gap to ML\@.
  Specifically, in the high SNR regime,
  MC requires at most $4/\MAR$ more measurements than lasso,
  where $\MAR$ is the mean-to-average ratio defined in (\ref{eq:MAR-def}).
  Both lasso and MC scale as $m = O(k\log(n-k))$.
  Thus, the benefit of lasso is not in its scaling with respect to the
  problem dimensions, but rather its ability to detect
  the sparsity pattern even in the presence
  of relatively small nonzero coefficients (i.e.\ low $\MAR$).
\end{itemize}

While our results settle the question of the optimal scaling of the number
of measurements $m$ in terms of $k$ and $n$, there is clearly
a gap in the necessary and sufficient conditions in terms of the scaling
of the SNR\@.  We have seen that full ML estimation could potentially have
a scaling in SNR as small as $m = O(1/\SNR) + k - 1$.  An open question
is whether there is any practical algorithm that can achieve a similar scaling.

A second open issue is to determine conditions for partial sparsity recovery.
The above results define conditions for recovering all the positions
in the sparsity pattern.  However, in many practical applications, obtaining
some large fraction of these positions would be sufficient.
Neither the limits of partial sparsity recovery nor
the performance of practical algorithms are completely understood,
though some results have been reported
in~\cite{AkcakayaT:CISS08,AkcakayaT:07arXiv,Reeves:08,AeronZS:07arXiv}.

\appendix

\subsection{Deterministic Necessary Condition}

The proof of Theorem~\ref{thm:minMeasML} is based on the following
deterministic necessary condition for sparsity recovery.
Recall the notation that for any $J \subseteq \{1,2,\ldots,n\}$,
$P_J$ denotes the orthogonal projection onto the span of the vectors
$\{a_j\}_{j\in J}$.
Additionally, let $P_J^\perp = I - P_J$
denote the orthogonal projection onto the orthogonal complement of
$\span(\{a_j\}_{j \in J})$.

\begin{lemma} \label{lem:projDiff}
A necessary condition for ML detection to succeed
(i.e.\ $\IhatML = \Itrue$) is:
\beq \label{eq:detIneq}
\mbox{for all $i \in \Itrue$ and $j \not \in \Itrue$},
\quad
    \frac{|a_i'P_K^\perp y|^2}{a_i'P_K^\perp a_i} \geq
    \frac{|a_j'P_K^\perp y|^2}{a_j'P_K^\perp a_j}
\eeq
where $K = \Itrue \setminus \{i\}$.
\end{lemma}
\begin{proof}
Note that $y = P_K y + P_K^\perp y$ is an orthogonal decomposition of $y$ into
the portions inside and outside the subspace $S = \span(\{a_j\}_{j\in K})$.
An approximation of $y$ in subspace $S$ leaves residual $P_K^\perp y$.
Intuitively, the condition (\ref{eq:detIneq})
requires that the residual be at least as highly correlated with the remaining
``correct'' vector $a_i$ as it is with any of the ``incorrect'' vectors
$\{a_j\}_{j\not\in\Itrue}$.

Fix any $i \in \Itrue$, $j \not\in \Itrue$ and let
\[
    J = K \cup \{j\} = (\Itrue \setminus \{i\}) \cup \{j\}.
\]
That is, $J$ is equal to the true sparsity pattern $\Itrue$,
except that a single ``correct'' index $i$ has been replaced by an
``incorrect'' index $j$.
If the ML estimator is to select $\IhatML = \Itrue$ then the
energy of the noisy vector $y$ must be larger on the true subspace $\Itrue$,
than the incorrect subspace $J$.  Therefore,
\beq \label{eq:projIneq}
    \|P_{\Itrue} y \|^2 \geq \|P_J y\|^2.
\eeq
Now, a simple application of the matrix inversion lemma shows that
since $\Itrue = K \cup \{i\}$,
\begin{subequations}
\begin{equation}
 \label{eq:lemma1a}
    \|P_{\Itrue} y\|^2 = \|P_Ky\|^2 + \frac{|a_i'P_K^\perp y|^2}{a_i'P_K^\perp a_i}.
\end{equation}
Also, since $J = K \cup \{j\}$,
\begin{equation}
 \label{eq:lemma1b}
    \|P_J y\|^2 = \|P_Ky\|^2 + \frac{|a_j'P_K^\perp y|^2}{a_j'P_K^\perp a_j}.
\end{equation}
\end{subequations}
Substituting (\ref{eq:lemma1a})--(\ref{eq:lemma1b}) into (\ref{eq:projIneq})
and cancelling $\|P_Ky\|^2$ shows (\ref{eq:detIneq}).
\end{proof}

\subsection{Proof of Theorem \ref{thm:minMeasML} }
\label{sec:NecProof}
To simplify notation, assume without loss of generality that
$\Itrue = \{1,2,\ldots,k\}$.
Also, assume that the minimization in (\ref{eq:MAR-def})
occurs at $j=1$ with
\beq \label{eq:xmin}
    |x_1|^2 = \frac{m}{k} \SNR \cdot \MAR.
\eeq
Finally, since adding measurements (i.e.\ increasing $m$) can only improve the
chances that ML detection will work, we can assume that in addition
to satisfying (\ref{eq:minMeasML}), the numbers of measurements satisfy
the lower bound
\beq \label{eq:measMinEpsilon}
    m > \epsilon k\log(n-k) + k -1,
\eeq
for some $\epsilon > 0$.
This assumption implies that
\beq \label{eq:measLim}
    \lim \frac{\log(n-k)}{m-k+1} =
    \lim \frac{1}{\epsilon k} = 0.
\eeq
Here and in the remainder of the proof the limits are as $m$, $n$ and
$k \arr \infty$ subject to (\ref{eq:minMeasML}) and (\ref{eq:measMinEpsilon}).
With these requirements on $m$,
we need to show $\lim \Pr(\IhatML = \Itrue) = 0$.

From Lemma \ref{lem:projDiff}, $\IhatML = \Itrue$ implies (\ref{eq:detIneq}).
Thus
\beqan
  \lefteqn{\Pr\left(\IhatML = \Itrue \right)} \\
   & \leq & \Pr\left(
      \frac{|a_i'P_K^\perp y|^2}{a_i'P_K^\perp a_i} \geq
      \frac{|a_j'P_K^\perp y|^2}{a_j'P_K^\perp a_j}
      \quad \mbox{$\forall$ $i \in \Itrue$, $j \not\in \Itrue$}
        \right) \\
   & \leq & \Pr\left(
      \frac{|a_1'P_K^\perp y|^2}{a_1'P_K^\perp a_1} \geq
      \frac{|a_j'P_K^\perp y|^2}{a_j'P_K^\perp a_j}
      \quad \mbox{$\forall$ $j \not\in \Itrue$}
        \right) \\
   & = & \Pr( \DeltaSub \geq \DeltaAdd),
\eeqan
where 
\beqan
    \DeltaSub &=&  \frac{1}{\log(n-k)}\frac{|a_1'P_K^\perp y|^2}{a_1'P_K^\perp a_1}, \\
    \DeltaAdd &=&  \frac{1}{\log(n-k)}\max_{j \in \{k+1,\ldots,n\}} \frac{|a_j'P_K^\perp y|^2}{a_j'P_K^\perp a_j},
\eeqan
and $K = \Itrue \setminus \{1\} = \{2,\ldots,k\}$.
The $-$ and $+$ superscripts
are used to reflect that $\DeltaSub$ is the energy lost from removing
``correct'' index 1,
and $\DeltaAdd$ is the energy added from adding the worst ``incorrect" index.
The theorem will be proven if we can show that
\beq \label{eq:deltaLim}
    \limsup \DeltaSub < \liminf \DeltaAdd
\eeq
with probability approaching one.
We will consider the two limits separately.

\subsubsection{Limit of $\DeltaAdd$}  Let $V_K$ be the $k-1$
dimensional space spanned by the vectors $\{a_j\}_{j\in K}$.
For each $j \not \in \Itrue$, let $u_j$ be the unit vector
\[
    u_j = P_K^\perp a_j / \|P_K^\perp a_j\|.
\]
Since $a_j$ has i.i.d.\ Gaussian components, it is spherically symmetric.
Also, if $j \not \in K$, $a_j$ is independent of the subspace $V_K$.
Hence, in this case, $u_j$ will be a unit vector uniformly distributed
on the unit sphere in $V_K^{\perp}$.
Since $V_K^{\perp}$ is an $m-k+1$ dimensional subspace,
it follows from Lemma~\ref{lem:betaProj}
(see Appendix~~\ref{sect:maxChiSq})
that if we define
\[
    z_j = |u_j'P_K^\perp y|^2 / \|P_K^\perp y\|^2,
\]
then $z_j$ follows a $\BetaDist(1,m-k+1)$ distribution.
See Appendix~\ref{sect:maxChiSq} for a review of the chi-squared
and beta distributions and some simple results on these variables
that will be used in the proofs below.

By the definition of $u_j$,
\[
    \frac{|a_j'P_K^\perp y|^2}{a_j'P_K^\perp a_j} = |u_j'P_K^{\perp}y|^2
        = z_j \|P_K^{\perp}y\|^2,
\]
and therefore
\beq \label{eq:deltaAdda}
    \DeltaAdd
     = \frac{1}{\log(n-k)} \|P_K^{\perp}y\|^2 \max_{j \in \{k+1,\ldots,n\}} z_j.
\eeq

Now the vectors $a_j$ are independent of one another, and
for $j \not \in \Itrue$, each $a_j$ is independent of $P_K^\perp y$.
Therefore, the variables $z_j$ will be i.i.d.  Hence,
using Lemma \ref{lem:betaMax}
(see Appendix~\ref{sect:maxChiSq})
and (\ref{eq:measLim}),
\beq \label{eq:zlim}
    \lim \frac{m-k+1}{\log(n-k)} \max_{j=k+1,\ldots,n} z_j = 2
\eeq
in distribution.
Also,
\beqa
    \liminf \frac{1}{m-k+1}\|P_K^{\perp}y\|^2
    &\stackrel{(a)}{\geq}& \lim \frac{1}{m-k+1}\|P_{\Itrue}^{\perp}y\|^2
        \nonumber \\
    &\stackrel{(b)}{=}&    \lim \frac{1}{m-k+1}\|P_{\Itrue}^{\perp}d\|^2
        \nonumber \\
    &\stackrel{(c)}{=}&    \lim \frac{m-k}{m-k+1} = 1
        \label{eq:pylim}
\eeqa
where (a) follows from the fact that $K \subset \Itrue$ and
hence $\|P_K^\perp y\| \geq \|P_{\Itrue}^\perp y\|$;
(b) is valid since $P_{\Itrue}^\perp a_j = 0$ for all $j \in \Itrue$
and therefore $P_{\Itrue} x = 0$;
and (c) follows from the fact that $P_{\Itrue}^\perp d$ is a unit-variance
white random vector in an $m-k$ dimensional space.
Combining (\ref{eq:deltaAdda}), (\ref{eq:zlim}) and (\ref{eq:pylim}) shows that
\beq \label{eq:deltaAddLim}
    \liminf \DeltaAdd \geq 2.
\eeq

\subsubsection{Limit of $\DeltaSub$}
For any $j \in K$, $P_K^\perp a_j = 0$.  Therefore,
\[
    P_K^\perp y =  P_K^\perp \left(\sum_{j=1}^k a_jx_j + d\right) =
    x_1 P_K^\perp a_1 + P_K^\perp d.
\]
Hence,
\[
    \frac{|a_1'P_K^\perp y|^2}{a_1'P_K^\perp a_1}
    = \left| \|P_K^\perp a_1\| x_1 + v \right|^2,
\]
where $v$ is given by
\[
    v = a_1'P_K^\perp d / \|P_K^\perp a_1\|.
\]
Since $P_K^\perp a_1 / \|P_K^\perp a_1\|$ is a random unit vector independent
of $d$, and $d$ is a zero-mean, unit-variance Gaussian random vector,
$v \sim {\mathcal{N}}(0,1)$.
Therefore,
\beqa
    \lim \DeltaSub &=&
    \lim \left| \frac{\|P_K^\perp a_1\|x_1}{\log^{1/2}(n-k)} +
    \frac{1}{\log^{1/2}(n-k)}v\right|^2 \nonumber \\
    &=& \lim \frac{\|P_K^\perp a_1\|^2|x_1|^2}{\log(n-k)},
    \label{eq:deltaSubLima}
\eeqa
where, in the last step, we used the fact that
$$v / \log^{1/2}(n-k) \arr 0.$$
Now, $a_1$ is a Gaussian vector with variance $1/m$ in each component and
$P_K^\perp$ is a projection onto an $m-k+1$ dimensional space.  Hence,
\beq \label{eq:pkaone}
    \lim \frac{m\|P_K^\perp a_1\|^2}{m-k+1} = 1.
\eeq
Starting with a combination of
(\ref{eq:deltaSubLima}) and (\ref{eq:pkaone}),
\beqa
    \limsup \DeltaSub
     & = & \limsup \frac{|x_1|^2(m-k+1)}{m \log(n-k)} \nonumber \\
     &\stackrel{(a)}{=}& \limsup \frac{(\SNR \cdot \MAR) (m-k+1)}{k\log(n-k)}
                                                                   \nonumber \\
     &\stackrel{(b)}{<}& 2 - \delta
\label{eq:deltaSubLim}
\eeqa
where (a) uses (\ref{eq:xmin}); and (b) uses (\ref{eq:minMeasML}).

Comparing (\ref{eq:deltaAddLim}) and (\ref{eq:deltaSubLim})
proves (\ref{eq:deltaLim}), thus completing the proof.

\subsection{Proof of Theorem \ref{thm:corrBnd}}
\label{sec:SuffProof}
We will show that there exists a $\mu > 0$ such that, with high probability,
\beq \label{eq:muCond}
    \begin{array}{cl}
        |a_i'y|^2 > \mu & \mbox{for all $i \in \Itrue$}; \\
        |a_j'y|^2 < \mu & \mbox{for all $j \not \in \Itrue$}.
    \end{array}
\eeq
When (\ref{eq:muCond}) is satisfied,
$$
 |a_i'y| > |a_j'y| \quad \mbox{for all indices $i \in \Itrue$
    and $j \not \in \Itrue$}.
$$
Thus, (\ref{eq:muCond}) implies that the maximum correlation estimator
$\IhatCorr$ in (\ref{eq:corrEst}) will select $\IhatCorr = \Itrue$.
Consequently, the theorem will be proven
if can find a $\mu$ such that (\ref{eq:muCond}) holds with high probability.

Since $\delta > 0$, we can find an $\epsilon > 0$ such that
\beq \label{eq:epsMC}
    \sqrt{8+\delta} - \sqrt{2+\epsilon} > \sqrt{2}.
\eeq
Define
\beq \label{eq:muDef}
    \mu = (2+\epsilon)(1+\SNR)\log(n-k).
\eeq
Define two probabilities corresponding to the two conditions in (\ref{eq:muCond}):
\beqa
    \PMD &=& \Pr\left( |a_i'y|^2 < \mu \quad \mbox{for some $i \in \Itrue$} \right)
        \label{eq:pmiss} \\
    \PFA &=& \Pr\left( |a_j'y|^2 > \mu \quad \mbox{for some $j \not \in \Itrue$} \right).
        \label{eq:pfa}
\eeqa
The first probability $\PMD$ is the probability of missed detection,
i.e., the probability that the energy on
one of the ``true'' vectors, $a_i$ with $i \in \Itrue$,
is below the threshold $\mu$.
The second probability $\PFA$ is the false alarm probability,
i.e., the probability that the energy on
one of the ``incorrect'' vectors, $a_j$ with $j \not \in \Itrue$,
is above the threshold $\mu$.
Since the correlation estimator detects the correct sparsity pattern
when there are no missed vectors or false alarms, we have the bound
\[
    \Pr\left(\IhatCorr \neq \Itrue\right) \leq \PMD + \PFA.
\]
So the result will be proven if we can show that
$\PMD$ and $\PFA$ approach zero as $m$, $n$ and $k \arr \infty$
satisfying (\ref{eq:minMeasCorr}).
We analyze these two probabilities separately.

\subsubsection{Limit of $\PFA$}
Consider any index $j \not \in \Itrue$.
Since $y$ is a linear combination of vectors $\{a_i\}_{i\in\Itrue}$
and the noise vector $d$, $a_j$ is independent of $y$.  Also, recall that
the components of $a_j$ are ${\mathcal{N}}(0,1/m)$.
Therefore, conditional on $\|y\|^2$, the inner product $a_j'y$ is
Gaussian with mean zero and variance $\|y\|^2 / m$.  For large $m$,
$\|y\|^2/m \arr 1 + \SNR$.  Hence, we can write
\[
    |a_j'y|^2 = (1+\SNR) u_j^2,
\]
where $u_j$ is a random variable that converges in distribution to a
zero mean Gaussian with unit variance.
Using the definitions of $\PFA$ in (\ref{eq:pfa}) and
$\mu$ in (\ref{eq:muDef}), we see that
\beqan
   \PFA &=& \Pr\left( \max_{j \not \in \Itrue} |a_j'y|^2 > \mu \right) \\
        &=& \Pr\left( \max_{j \not \in \Itrue} (1+\SNR)u_j^2 > \mu \right) \\
        &=& \Pr\left( \max_{j \not \in \Itrue} u_j^2 > \mu/(1+\SNR) \right) \\
        &=& \Pr\left( \max_{j \not \in \Itrue} u_j^2 > (2+\epsilon)\log(n-k) \right)
            \arr 0
\eeqan
where the last limit uses
Lemma~\ref{lem:maxChiSq}
(see Appendix~\ref{sect:maxChiSq})
on the maxima of chi-squared random variables.

\subsubsection{Limit of $\PMD$}
Consider any index $i \in \Itrue$.  Observe that
\[
    a_i'y = \|a_i\|^2|x_i|^2 + a_i'e_i,
\]
where
\[
    e_i = y - a_ix_i = \sum_{\ell \in \Itrue, \ell \neq i} a_\ell x_\ell + d.
\]
It is easily verified that $\|a_i\|^2 \arr 1$ and $\|e_i\|^2/m \arr 1+\SNR$.
Using a similar argument as above, one can show that
\beq \label{eq:axu}
    a_i'y = x_i + (1+\SNR)^{1/2}u_i,
\eeq
where $u_i$ approaches a zero-mean, unit-variance Gaussian in distribution.

Now, using (\ref{eq:snrVal}), (\ref{eq:MAR-def}) and (\ref{eq:minMeasCorr}),
\beqa
    |x_i|^2 &\geq& \frac{\MAR \|x\|^2}{k} = \frac{m \, \MAR \cdot \SNR}{k}
        \nonumber\\
        &>&  (8+\delta)(1+\SNR)\log(n-k).\label{eq:xlb}
\eeqa
Combining (\ref{eq:epsMC}), (\ref{eq:muDef}), (\ref{eq:axu}) and
(\ref{eq:xlb})
\beqan
    |a_i'y|^2 \leq \mu & \Longleftrightarrow &
    \left|~ x_i + (1+\SNR)^{1/2}u_i \right| \leq \mu^{1/2}  \\
    & \Longrightarrow &(1+\SNR)u_i^2 \geq \left(|x_i| - \mu^{1/2}\right)^2  \\
    & \Longleftrightarrow & u_i^2 > 2 \log(n-k) \\
    & \Longrightarrow & u_i^2 > 2 \log(k)
\eeqan
where, in the last step,
we have used the fact that since $k/n < 1/2$, $n-k > k$.
Therefore, using Lemma \ref{lem:maxChiSq} 
\beqa
    \PMD &=& \Pr\left( \min_{i \in \Itrue} |a_i'y|^2 \leq \mu\right)
        \nonumber \\
    &\leq & \Pr\left( \max_{i \in \Itrue} u_i^2 > 2 \log(k)\right)
    \arr 0.
\eeqa
Hence, we have shown both $\PFA \arr 0$ and $\PMD \arr 0$ as
$n \arr \infty$, and the theorem is proven.

\subsection{Maxima of Chi-Squared and Beta Random Variables}
\label{sect:maxChiSq}

The proofs of the main results above require a few simple results
on the maxima of large numbers of chi-squared and beta random variables.
A complete description of chi-squared and beta random variables
can be found in \cite{EvansHP:00}.

A random variable $U$ has a \emph{chi-squared} distribution with $r$
degrees of freedom if it can be written as
\[
    U = \sum_{i=1}^r Z_i^2,
\]
where $Z_i$ are i.i.d.\ ${\mathcal{N}}(0,1)$.
For every $n$ and $r$ define the random variables
\beqan
    \Mover_{n,r}  &=& \max_{i\in\{1,\ldots,n\}} U_i, \\
    \Munder_{n,r} &=& \min_{i\in\{1,\ldots,n\}} U_i,
\eeqan
where the $U_i$'s are i.i.d.\ chi-squared with $r$ degrees of freedom.

\begin{lemma} \label{lem:maxChiSqLog}
For $\overline{M}_{n,r}$ defined as above,
\[
    \lim_{n \arr \infty} \frac{1}{\log(n)} \Mover_{n,1} = 2,
\]
where the convergence is in distribution.
\end{lemma}
\begin{proof}
We can write $\Mover_{n,1} = \max_{i\in\{1,\ldots,n\}} Z_i^2$
where $Z_i$ are i.i.d.\ ${\mathcal{N}}(0,1)$.
Then, for any $a > 0$,
\beqan
 \lefteqn{\Pr\left( \frac{1}{\log(n)}\Mover_{n,1} <  a \right)} \\
   &=& \Pr\left( |Z_1|^2 <  a \log(n) \right)^n \\
   &=& \mbox{erf}\left(\sqrt{a\log(n)/2}\right)^n  \\
   &\approx& \left[1-\sqrt{\frac{2}{\pi a\log(n)}}\exp(-a \log(n)/2)\right]^n \\
   &=& \left[1-\sqrt{\frac{2}{\pi a\log(n)}}\frac{1}{n^{a/2}}\right]^n
\eeqan
where the approximation is valid for large $n$.
Taking the limit as $n \arr \infty$, one can now easily show that
\[
    \lim_{n \arr \infty} \Pr\left( \frac{1}{\log(n)}\Mover_n <  a \right)
    = \left\{ \begin{array}{ll}
    0, & \mbox{for $a < 2$}; \\
    1, & \mbox{for $a > 2$}
    \end{array} \right.
\]
and therefore $\Mover_n/\log(n) \arr 2$ in distribution.
\end{proof}

\begin{lemma} \label{lem:maxChiSq}
In any limit where $r \arr \infty$ and $\log(n)/r \arr 0$,
\[
    \lim_{r \arr \infty} \frac{1}{r} \Mover_{n,r}
    = \lim_{r \arr \infty} \frac{1}{r} \Munder_{n,r} = 1,
\]
where the convergence is in distribution.
\end{lemma}
\begin{proof}
It suffices to show
\beqan
    \limsup_{r \arr \infty} \frac{1}{r} \Mover_{n,r} &\leq& 1, \\
    \liminf_{r \arr \infty} \frac{1}{r} \Munder_{n,r} &\geq& 1.
\eeqan
We will just prove the first inequality since the proof of the second
is similar.
We can write
\[
    \frac{1}{r} \Mover_{n,r} = \max_{i=1,\ldots,n} V_i,
\]
where each $V_i = U_i/r$ and the $U_i$'s are i.i.d.\
chi-squared random variables with $r$ degree of freedom.
Using the characteristic function of $U_i$ and Chebyshev's inequality, one
can show that for all $\epsilon > 0$,
\beqan
    \Pr( V_i > (1+\epsilon) ) &=& \Pr( U_i > (1+\epsilon)r ) \\
    &\leq& (1+\epsilon)e^{-\epsilon r/2}.
\eeqan
Therefore,
\beqan
    \Pr\left( \Mover_{n,r} \leq 1 + \epsilon \right)
    &=& \left[\Pr( V_i \leq 1 + \epsilon ) \right]^n \\
    &\geq& \left[1 -  (1+\epsilon)e^{-\epsilon r/2}\right]^n \\
    &\geq& 1 -  (1+\epsilon)ne^{-\epsilon r/2}\\
    &=& 1 -  (1+\epsilon)\exp\left[\log(n)-\epsilon r/2\right] \\
    &\arr& 1,
\eeqan
where the limit in the last step follows from the fact that $\log(n)/r \arr 0$.
Since this is true for all $\epsilon$ it follows that $\limsup r^{-1}\Mover_{n,r} \leq 1$.
Similarly, one can show $\liminf r^{-1}\Mover_{n,r} \geq 1$ and this proves the lemma.
\end{proof}

The next two lemmas concern certain beta distributed random variables.
A real-valued scalar random variable $W$ follows
a $\BetaDist(r,s)$ distribution if it can be written as
\[
    W = U_r / (U_r + V_s),
\]
where the variables $U_r$ and $V_s$ are independent chi-squared random variables
with $r$ and $s$ degrees of freedom, respectively.  The importance of the
beta distribution is given by the following lemma.

\begin{lemma} \label{lem:betaProj}
Let $x$ and $u \in \R^s$ be any two independent random vectors,
with $u$ being uniformly distributed on the unit sphere.
Let $w = |u'x|^2 / \|x\|^2$ be the energy of $w$ projected onto $u$.
Then $w$ is independent of $x$ and follows a $\BetaDist(1,s-1)$ distribution.
\end{lemma}
\begin{proof}
This can be proven along the lines of the arguments in
\cite{FletcherRGR:06}.
\end{proof}

The following lemma provides a simple expression for the
maxima of certain beta distributed variables.

\begin{lemma} \label{lem:betaMax}
Given any $s$ and $n$, let $w_{j,s}$, $j=1,\ldots,n$,
be i.i.d.\ $\BetaDist(1,s-1)$ random variables and define
\[
    T_{n,s} = \max_{j=1,\ldots,n} w_{j,s}.
\]
Then for any limit with $n$ and $s \arr \infty$ and
$\log(n) / s \arr 0$,
\[
    \lim_{n,s \arr \infty} \frac{s}{\log(n)}T_{n,s} = 2,
\]
where the convergence is in distribution.
\end{lemma}
\begin{proof}
We can write $w_{j,s} = u_j / (u_j + v_{j,s-1})$ where $u_j$ and $v_{j,s-1}$
are independent chi-squared random variables with 1 and $s-1$ degrees of
freedom, respectively.
Let
\beqan
    \Mover_n        &=& \max_{j\in\{1,\ldots,n\}} u_j \\
    \Mover_{n,s-1}  &=& \max_{j\in\{1,\ldots,n\}} v_{j,s-1} \\
    \Munder_{n,s-1} &=& \min_{j\in\{1,\ldots,n\}} v_{j,s-1}.
\eeqan
Using the definition of $T_{n,s}$,
\[
    T_{n,s} \leq \frac{ \Mover_n }{ \Mover_n + \Munder_{n,s-1}}.
\]
Now Lemmas~\ref{lem:maxChiSqLog} and~\ref{lem:maxChiSq} and the hypothesis
of this lemma show that $\Mover_n / \log(n) \arr 2$,
$\Munder_{n,s-1} / (s-1) \arr 1$, and $\log(n)/s \arr 0$.
One can combine these limits to show that
\[
    \limsup_{n,s  \arr \infty} \frac{s}{\log(n)}T_{n,s} \leq 2.
\]
Similarly, one can show that
\[
    \liminf_{n,s  \arr \infty} \frac{s}{\log(n)}T_{n,s} \geq 2,
\]
and therefore $s T_{n,s} / \log(n) \arr 2$.
\end{proof}

\section*{Acknowledgment}
The authors thank Martin Vetterli for his support, wisdom, and encouragement.

\bibliographystyle{IEEEtran}
\bibliography{bibl}

\end{document}